\documentclass[12pt]{article}
\usepackage[english]{babel}
\usepackage[normalem]{ulem}
\usepackage[all]{xy}
\usepackage{amsmath,amsthm,booktabs,color,dsfont,epsfig,graphicx,mathrsfs,multirow,scalefnt}
\usepackage{amsfonts, amssymb}
\usepackage{float}
\usepackage{fancyhdr}

\usepackage{mathtools}
\newcommand{\A}{\mathfrak{L}}

\newcommand\ml{\left\{\begin{array}}
\newcommand\mr{\end{array}\right\}}

\begin{document}

\pagestyle{fancy}

\title{Intra-strand symmetries and asymmetries in bacterial DNA: Evolutive features or relics of primordial genomes?}

\lhead{M Sobottka}
\rhead{Intra-strand symmetries and asymmetries in bacterial DNA}

\author{
\small{Marcelo Sobottka}\\
\footnotesize{UFSC -- Department of Mathematics}\\
\footnotesize{88040-900 Florian\'{o}polis - SC, Brazil}\\
\footnotesize{\texttt{marcelo.sobottka@ufsc.br}}
}

\date{\ }
\maketitle

\begin{abstract}
In this work we analyze some models used to explain the origins of intra-strand parity and strand compositional asymmetries in bacterial genomes. Due to the particular way that these two features emerge in bacterial DNA, we  performed our analysis from the perspective that they are complementary phenomena that should be addressed together. Although most of the models for these features try to explain them as consequence of evolutionary mechanisms, recently it was proposed that they could be `relics' of some primordial genome that were conserved thorough out the genome evolution. We shall pay special attention to the S-H model, which is, up to the date, the unique model proposed as a possible explanation for intra-strand parity and strand compositional asymmetries in primordial genomes as mere consequence of randomness under chemical/physical constraints. In particular, we shall discuss possible directions to test some of the hypotheses of the  S-H model, and we will present a possible formulation of the S-H model as an evolutive model too.
\end{abstract}

{\bf Keywords:} Genetics; Bacterial DNA; Intra-Strand Parity; Strand Compositional Asymmetry; Chargaff's Second Parity Rule


\section{Introduction}\label{sec:Introduction}

A genome is a duplex of DNA strands, one of them referred as the Watson strand and the other referred as the Crick strand in homage to the molecular biologists James Dewey Watson and Francis Harry Compton Crick, who unveiled the double helix structure of the DNA. Each of the strands consists of a sequence of nucleotide bases, that could be of four types: adenine ($A$), cytosine ($C$), guanine ($G$) and thymine ($T$). The strands are complementary one to the other in the sense that an adenine in one strand is paired with a thymine in the other strand, while a cytosine in one strand is paired with a guanine in the other strand. For sake of notation, let $\alpha:\{A,C,G,T\}\to\{A,C,G,T\}$ be the base pairing rule map (BPR map), that is, the map  which takes each nucleotide base to its complementary one. Due to the chemical composition of DNA molecules each strand can only be elongated in one direction ($5'\to3'$). Furthermore, the strands elongated in opposite directions.

The DNA molecules are constituted of three main structures: The primary structure concerning on the distribution of nucleotides along DNA strands; The secondary structure concerning on the base paring between the two strands; The tertiary structure concerning on the three dimensional shape of the DNA molecule.

Several underlying patterns in the primary structure of DNA have been observed \cite{Cattani12,CattaniPierro13,Chargaff,Eisen_et_Al_00,Li_92,Lobry96-2,Mackiewicz01,SobottkaHart11} and have lead scientists to propose models for evolutive pressures and mutational mechanisms acting on the genomes. In this article we shall concern on two distributional patterns observed in the primary structure, namely, the intra-strand parity and the strand compositional asymmetry.

Intra-strand symmetries in DNA sequences were early noticed by Chargaff \cite{Chargaff,ForsdykeMortimer,Rudner_et_al} who observed that in a same strand each nucleotide occurs with approximately the same frequency than its complementary nucleotide. Such characteristic is known as intra-strand parity (ISP) or Chargaff's second parity rule, and has been successfully tested for many double-stranded DNA genomes \cite{HartMartinez11,MitchellBridge}. Such property seems to be extensible for short oligonucleotides: The frequency of any short nucleotide sequence $i_1i_2...i_k$ on a strand is approximately equal to the frequency of its reverse complement $\alpha(i_k)...\alpha(i_2)\alpha(i_1)$ on the same strand \cite{Baisnee02,Forsdyke95,HartMartinez11,Okamura_et_al07,Prabhu93}.

On the other hand, strand compositional asymmetry (SCA) was noticed in viral and bacterial genomes \cite{FrancinoOchman97,FrankLobry99,GuoNing11,Lobry96-1,Mackiewicz04,MrazekKarlin98,Poptsova_et_al09,Rocha04,TillierCollins00_1,TillierCollins00_2}.  Indeed each strand in most of bacterial genomes presents a transition point before which it is possible to observe accumulation of one nucleotide compared to its complementary, and after which the same nucleotide is reduced with the same tax that it accumulated \cite{Poptsova_et_al09}. These two components of the strand were named chirochores \cite{Lobry96-1}, and in eubacteria, the transition point is related to the origin of replication and the chirochores coincide almost entirely with the replichores\footnote{The replication of the DNA duplex occurs by first unzipping the duplex into two single strands at some specific point named replication origin. From this point each strand is copied according to its own read direction until reach the point named replication terminus. A replichore is the portion of the strand between the origin and the terminus of the replication.} \cite{FrankLobry00}. Although this phenomenon is commonly observable in eubacteria where in general there is a single origin of replication, it was also noticed in viruses, archaea, mitochondria and parasites \cite{Grigoriev99,Lopez_et_al99,Miller_et_al03,Mohr_et_al99,MrazekKarlin98,NilssonAndersson05,Reyes_et_al98}.

\begin{center}
\begin{figure}
\begin{center}
\begin{minipage}[b]{.7\linewidth}
\includegraphics[width=\textwidth]{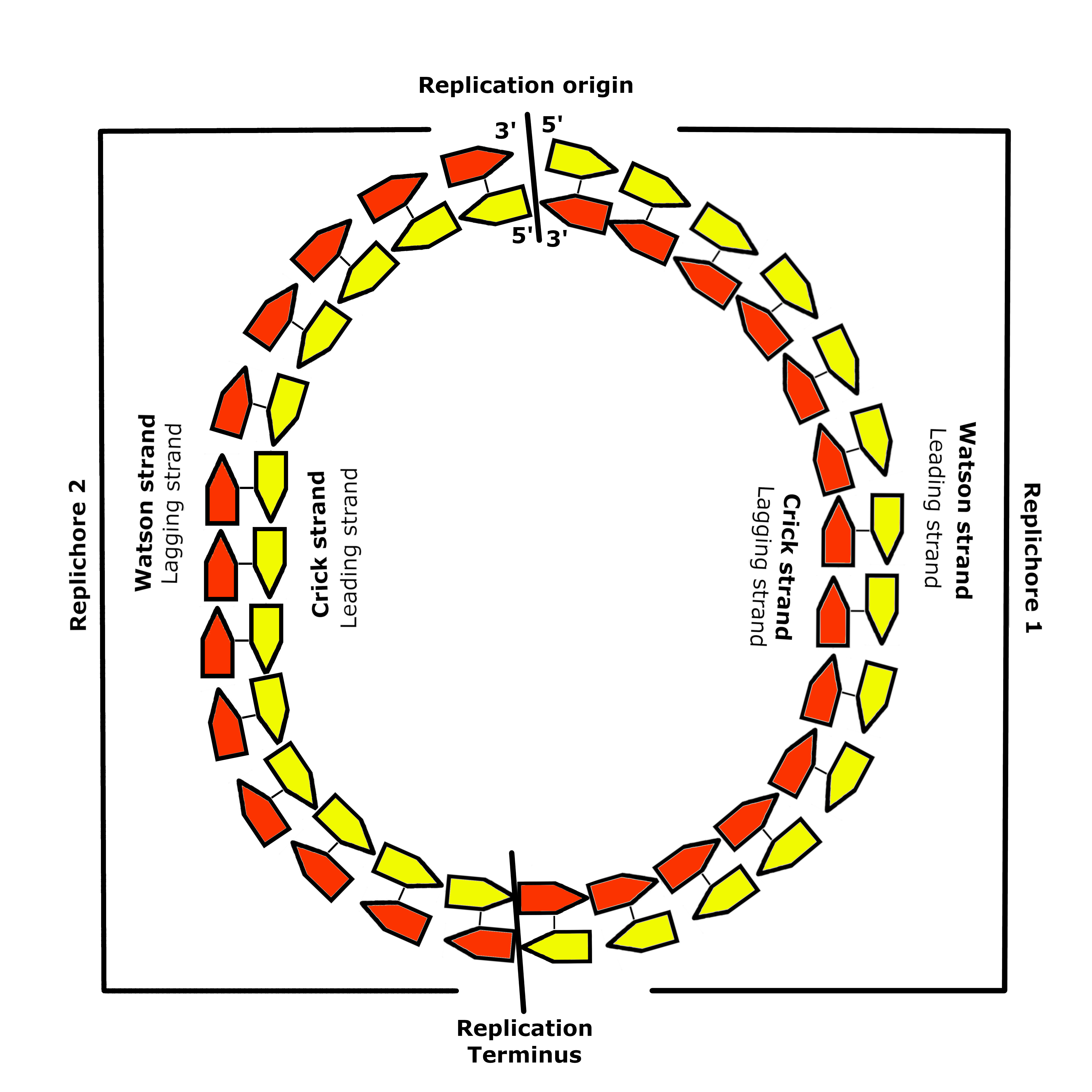}
\end{minipage}

\begin{minipage}[b]{.5\linewidth}
\includegraphics[width=\textwidth]{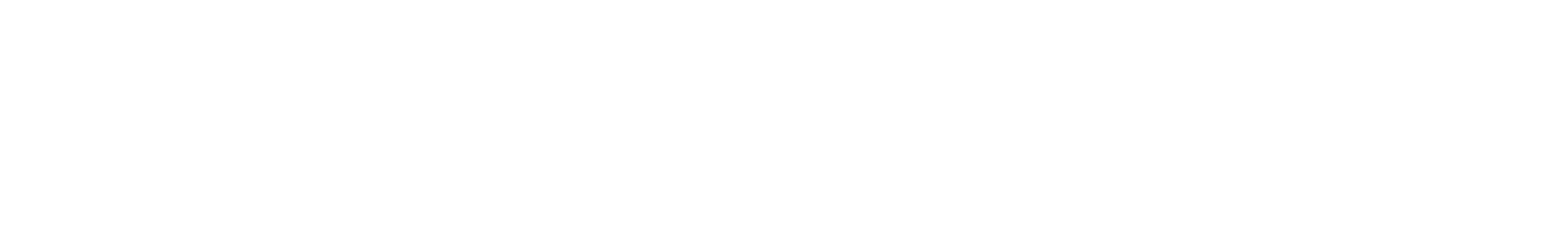}
\end{minipage}

\begin{minipage}[b]{.9\linewidth}
\includegraphics[width=\textwidth]{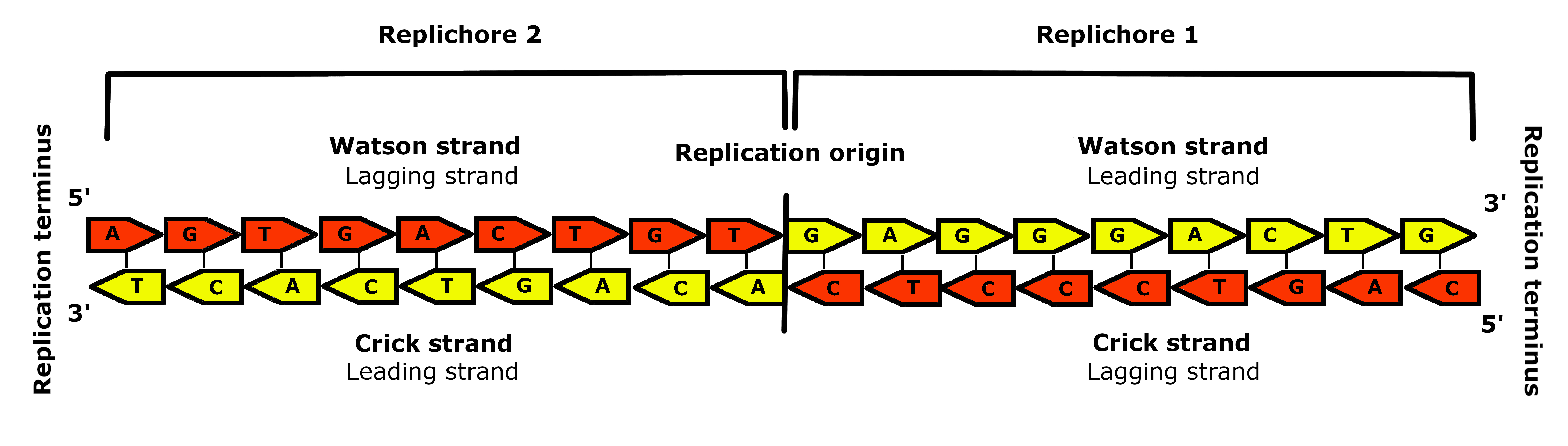}
\end{minipage}
\caption{At the top a schematic representation of a circular bacterial DNA sequence. At the bottom a schematic representation of the same bacterial DNA as a straight line. The direction $5'$ to $3'$ in each strand is related to the direction in which new nucleotide bases are added during the replication process.
It has been observed that in most of bacterial genomes, the lagging strand and the leading strand present different skews in the distribution of oligonucleotides (which is known as SCA). Since such skews neutralize mutually, in each whole strand the frequency of some oligonucleotide is almost the same than the frequency of their reverse complement (which is known as ISP). In particular, SCA and ISP together mean that the leading strands share a same nucleotide distribution, while the lagging strands share other nucleotide distribution.}\label{fig:genome}
\end{center}
\end{figure}
\end{center}

In this working we shall consider genomes of eubacteria. Eubacteria usually present circular genomes, with a single replication origin and replication terminus, that divide duplex into two replichores \cite{Lobry96-1} (it is very common that each replichore corresponds to about 50\% of the genome). We will represent such genomes either as a circle or as straight line (see Figure \ref{fig:genome}). In the circle representation, the north pole corresponds to the replication origin while the south pole corresponds to the replication terminus. The straight line presentation is thought as if we cut the south polo of the circle, and then the middle point of the line corresponds to the replication origin and both extremes of the line correspond to the replication terminus. In both presentations, the half of the DNA to the right of the replication origin will be considered the replichore 1, while the half of the DNA to the left of the replication origin will be considered the replichore 2. Finally, the Watson strand will be the external strand in the circle presentation and the top strand in the line presentation, while the Crick strand will be the internal strand in the circle presentation and the bottom strand in the line presentation. According to this convention replichore 1 of Watson strand and replichore 2 of the Crick strand are the leading strands during the replication, while replichore 2 of Watson strand and replichore 1 of the Crick strand are the lagging strands during the replication. For more details about the replication process we refer the reader to \cite{Alberts2002}.

Evolutive pressures acting on successive random mutations such as nucleotide
deletions, substitutions, insertions, duplications, inversions and inverted transpositions
\cite{Baldi2000,Bell97,Felsenstein81,Jukes69,Kimura80,Tavare89,Whittaker_et_al03} can efficiently explain most of the features in the genome structure. However, though substitutions, inversions and inverted transpositions are thought to be the cause of the ISP and SCA in the bacterial DNA \cite{AlbrechtBuehler2006,FrankLobry99,Jukes69,Lobry95,TillierCollins00_1}, the complete comprehension of these phenomena are likely far from being achieved. Recently, Zhang and Huang proposed that intra-strand symmetries and asymmetries could be relics of a primordial genome \cite{Koonin03} which were preserved along  evolution in spite of the mutations \cite{ZhangHuang08,ZhangHuang10-2}. A first model for the way as such characteristics could appear in primordial genomes is the  S-H model which was proposed in \cite{SobottkaHart11}. In the next sections we shall analyze some of the most accepted evolutive models and the S-H model. In particular, we will show that it is possible to formulate the S-H model as an evolutive model too, in a way compatible with biological mechanisms. Finally, we will use the S-H model to conjecture novel possible patterns shared by bacterial genomes.

We remark that the focus of this work is exclusively the analysis of the mathematical methods that has been used, and not to discuss or to propose biological mechanisms to validate such models. We hope that, by pointing out the strengths and weaknesses of mathematical explanations for ISP and SCA, we could help biologists in identifying the possible causes of these phenomena.

\section{Evolutive models}\label{EvolutiveModels}

Most of the proposed models for intra-strand symmetries and asymmetries are based on evolutive pressures and random mutations. In what follows we shall examine some of  such model and discuss how each of them fits for ISP and SCA.

\subsection{Substitution models}

A substitution model for nucleotide bases is given by the ordinary differential equation \begin{equation}\label{X'=MX}X'=MX\end{equation} where $X=X(t)=(X_{A}(t),X_{ C}(t),X_{G}(t),X_{T}(t))$ represents the quantity\footnote{Many times, instead to consider the total quantity of each nucleotide type in the strand, it is considered the frequency (or probability) of the nucleotide type in the strand.}  of each nucleotide in the strand at time $t$, and $M$ is matrix in the form
\begin{equation}\label{edoM}M=\left(\begin{array}{cccc}-m_A & m_{CA} & m_{GA} & m_{TA}\\\\
                                              m_{AC} & -m_{C} & m_{GC} & m_{TC}\\\\
                                              m_{AG} & m_{CG} & -m_{G} & m_{TG}\\\\
                                              m_{AT} & m_{CT} & m_{GT} & -m_{T}\\\\
                                \end{array}\right),\end{equation}
where $m_i,m_{ij}>0$ for all $i,j=A,C,G,T$ and the sum over each column is zero. Hence for each $i,j=A,C,G,T$ the quantity $-m_i$ is the reduction rate of the nucleotide type $i$ in the strand, and $m_{ij}$ is rate with a nucleotide of type $i$ is replaced for a nucleotide of type $j$.

Several substitution models have been proposed by varying the matrix $M$ or by considering additional hypotheses on how the substitutions rates vary on the strands (see \cite{Felsenstein81,Hasegawa_et_al85,Jukes69,Kimura80,Kimura81,Tamura92,Tamura93,Tavare86}).
We remark that substitution models were not originally thought for explaining ISP or any other specific pattern of the genomes, but with the aim of studying the genome evolution and as tools to estimate the evolutionary distance between two given genomes. Some of them even assume ISP as an {\em a priori} feature of the genomes \cite{Tamura92}. 

In \cite{Lobry95} Lobry observed that due to the Watson-Crick pairing in the DNA double helix, if a nucleotide of type $i$ is replaced for a nucleotide of type $j$ in one strand, then in the other strand a nucleotide of type $\alpha(i)$ is replaced for a nucleotide of type complementary to $\alpha(j)$. Hence, if one assumes {\em no-strand-bias conditions}, that is, if one assumes that the rates $m_i$ and $m_{ij}$ are the same in both strands, it follows that $m_{ij}=m_{\alpha(i)\alpha(j)}$. Therefore,  Lobry proved that \eqref{X'=MX} has an asymptotically stable equilibrium point $(\bar X_{A},\bar X_{ C},\bar X_{G},\bar X_{T})$ with $\bar X_{A}=\bar X_{ T}$ and $\bar X_{C}=\bar X_{G}$. Observe that the model supposes that the probability of a substitution occurring is independently on the site and that the substitutions are independent random events. Therefore, one can conclude that the final genome would present ISP for oligonucleotides.

In other words, Lobry provided a mathematical evidence that substitutions together the Watson-Crick pairing could be the cause of ISP. 
However, SCA consists in the most important hindrance for using substitution models to derive ISP in eubacteria (in particular for using substitution models with {\em no-strand-bias conditions}), since whenever we assume that substitutions occur homogeneously along both strands, we could restrict the process to any portion of the genome and we will find the same equilibrium point holding ISP, which means that SCA would not emerge. Thus, it is likely that substitutions where {\em no-strand-bias conditions} hold (at least in a large portion of the genome) could have strongly contributed for ISP only in genomes where SCA does not hold (and perhaps SCA was eliminated from those genomes due to such substitutions), but not in bacterial genomes.\\

On the other hand, ISP could emerge together SCA when the {\em no-strand-bias conditions} for substitutions are dismissed. In fact, among many underlying mutational- and selective mechanism-based hypotheses for the SCA, one of the most accepted is that SCA arises from the fact that the DNA replication has different mechanisms in the leading strand and in the lagging strand (see \cite{FrankLobry99,Rocha04} for extensive reviews on the subject). Such difference between leading and lagging strands would lead for different substitution rates in each of them.  

In \cite{FrankLobry99} Frank and Lobry analyzed several explanations that could explain why the leading strand in most of bacterial genomes present an excess of keto bases (G+T) over amino bases (A+C). All mechanisms analyzed are based on: $i.$ Strand bias causing different substitution rates in leading and lagging strands, making amino bases more likely to be replaced for keto bases in the leading strand than in the lagging strand (the deamination of the leading strand); The Watson-Crick pairing of the leading strand in one strand with the lagging strand in the other strand  causes that the excess of keto bases over amino bases in leading strands would be similar to the excess of amino bases over keto bases in lagging strands. Hence, by supposing that both leading strands are submitted to the same evolutive pressures, and still hold $m_i,m_{ij}>0$ for all $i,j=A,C,G,T$, it follows that they will reach the same skews of keto bases over amino bases, while both lagging strands will present similar rate skews, but of amino bases over keto bases. That is, the replichores will become chirochores and each whole strand will present ISP. We remark that this is a longtime process which take several generations to converge for such configuration. In fact, starting with a first genome, such process would occur during the replication and at the end we will have two organism, each of one carrying the original leading strand that suffered deamination and other which was replicated from the lagging strand used as template, and then without deamination. However, after several replications where the leading strands suffer deamination, most of the descendants will present similar deamination in both leading strands.

Note that, by focusing only on the deamination of the leading strands, this process does not explain SCA completely, since tit could not explain why it is observed an excess of purine bases (A+G) over pyrimidine bases (C+T) in the leading strand of bacterial genomes \cite{Lobry96-1}. 
However, although there is still not a widely accepted biological mechanism to explain the observed depyrimidination of the leading strands, if one assumes that the pyrimidine bases have also more chance to be replaced for purine bases in the leading strand, then we would obtain SCA as seen in actual genomes\footnote{In \cite{FrankLobry99} the authors discuss the purine-pyrimidine skew of the leading strand, but did not include a depyrimidination of the leading strand in the process due to lack of evidence of a biological mechanism for this. However here we are interested only on the possible stochastic processes that could lead for ISP and SCA without addressing biological mechanism.}.\\

Observe that much the substitutions proposed by Lobry to explain ISP in \cite{Lobry95} as the substitutions analyzed by Frank and Lobry in \cite{FrankLobry99} to explain ISP and the keto-amino SCA are strongly linked to the Watson-Crick pairing of the DNA double helix.

\subsection{Inverted transpositions}

Other mutational mechanisms acting on genomes are the inversions and inverted transpositions. An inversion is the process where some oligonucleotide changes its orientation but remain in the same strand. An inverted transposition is the process where some oligonucleotide changes its orientation and change from one strand to the other. Inversions and inverted transpositions was proposed by Albrecht-Buehler \cite{AlbrechtBuehler2006} to explain ISP. 

Observe that an inversion does not change the frequency of nucleotides in the strands. Therefore, if
we assume that some genome does not present ISP and SCA, then inversions would not generate these features. On the other hand, inverted transposition that change an oligonucleotide $i_1i_2...i_k$ from the Watson strand to Crick strand is also changing the oligonucleotide $\alpha(i_k)...\alpha(i_2)\alpha(i_1)$ from the Crick strand to Watson strand. Hence, if several inverted transpositions take place, then the process should converge to the situation where the Watson strand and the Crick strand have the same distribution of nucleotides, and both strands should present ISP for oligonucleotides. This is because inverted transpositions always diminishes the frequency of some oligonucleotide in one strand and increases the frequency of its reverse complementary, and since oligonucleotides that are more abundant in the strand have more chance to be randomly selected for being inverted and transposed, the process should converge to equilibrium. 

Once again, we see that the Watson-Crick pairing is underlying the possible explanation for ISP. However, the process as described above, without additional constraints does not lead necessarily to SCA. To derive SCA from inverted transpositions
we should also suppose that the initial genome has different skews in the nucleotide distributions of replichore 1 and replichore 2 (but not necessarily SCA). Furthermore, inverted transpositions should occur predominantly by changing oligonucleotides from one replichore to the other replichore, that is, by moving oligonucleotides from the leading (lagging) strand of the Watson strand to the leading (lagging) strand of the Crick strand. Thus, as the time goes, it is expected that the leading strands of the Watson strand and of the Crick strand have the same distribution of oligonucleotides (and the same for the lagging strands of both strands).  Under these assumption, inverted transpositions could explain ISP. However, this process would generate SCA only partially, since without other constraints it does not lead necessarily to the keto-amino and purine-pyrimidine skews that have been observed in the leading strands of actual bacterial genomes.

\section{Conserved patterns}

The existence of underlying patterns shared by many DNA sequences was first observed by Rogerson \cite{Rogerson89,Rogerson91}, who analyzed tetranucleotide frequencies of 18 viral and bacterial genomes and found out a statistically similar distribution pattern as in coding as in non-coding regions.  Later, Zhang et al. \cite{Zhang2013} analyzed the dinucleotide frequencies of 1300 species of archaea and bacteria found out that the frequencies of dinucleotides $AC$, $AG$, $CA$, $CT$, $GA$, $GT$, $TC$ and $TG$ present small deviation of their own means. That is, the frequencies of these dinucleotides do not vary much in the examined genomes, indicating some patterns which are shared by many distinct species. Other shared underlying patterns was found by Sobottka and Hart \cite{SobottkaHart10,SobottkaHart11} who examined 1049 bacterial genomes: If one plots on a same graphic the relationship between the frequency of any fixed mononucleotide and the frequency of any fixed dinucleotide, she will find out that the points seem to be clustered around a curve (see \cite[Figure 1]{SobottkaHart10} and \cite[Fig. 2]{SobottkaHart11}).

Note that, distinctly of ISP and SCA which are patterns that appear into the genome of a given bacteria, all the above mentioned patterns are not observed into a single genome, but observed when we consider a large set of distinct genomes, that is, they are underlying patterns of the domain Bacteria. The discovery of shared patterns leads Rogerson to propose that such patterns are not related to specific DNA functions and would have been conserved through evolution \cite{Rogerson91}. \\

In \cite{ZhangHuang08,ZhangHuang10-2} Zhang and Huang presented a theoretical basis for considering ISP an evolutionary conserved pattern (a relic of a primordial genome), and in \cite{SobottkaHart11} it was proposed a stochastic process to construct sequences of nucleotides presenting ISP and SCA in a very close way than bacterial genomes present them. The so called S-H model uses a fixed matrix \begin{equation}\label{aleph}\A=(L_{ij})_{i,j=A,C,G,T},\qquad where\ 0<L_{ij}=L_{\alpha(j)\alpha(i)}<1\end{equation} and a fixed probability vector  \begin{equation}\label{probM}M=(M_j)_{j=A,C,G,T}\qquad where\ 0<M_j=M_{\alpha(j)}<1\end{equation} Therefore, a duplex is constructed as follows:

\begin{description}
\item[Step 0] Start with some sequence of DNA duplex (possibly a single nucleotide duplex) \begin{equation}\label{eq:primitiveDNA}\left(\begin{array}{cccc} x_0 & x_1 & ... & x_k\\y_0 & y_{-1} & ... & y_{-k}\end{array}\right),\end{equation} which for sake of convention we suppose that the $x_0x_1...x_k$ is the proto-Watson strand and its elongation occurs from the left ($5'$) to the right ($3'$), while $y_{-k}...y_{-1}y_0=\alpha(x_{k})...\alpha(x_{1})\alpha(x_0)$ is the proto-Crick strand and its elongation is from the right ($5'$) to the left ($3'$).

\item[Step 1] After some stochastic time a new nucleotide duplex $\binom{\bar x}{\alpha(\bar x)}$ is randomly selected with some probability $M_{\bar x}=M_{\alpha(\bar x)}$.

\item[Step 2] With probability 1/2 the nucleotide $\binom{\bar x}{\alpha(\bar x)}$ approaches the original nucleotide from the right side of the original sequence, and with probability 1/2 it approaches the original sequence from the left side.

\item[step 3] If  $\binom{\bar x}{\alpha(\bar x)}$ approaches from the right side, then it has probability $L_{x_k\bar x}=L_{\alpha(\bar x)\alpha(x_k)}$ of to be accepted and to be the next nucleotide in the sequence. If it is accepted, then it is denoted as $\binom{ x_{k+1}}{y_{-k-1}}$ and the sequence becomes $$\left(\begin{array}{ccccc} x_0 & x_1 & ... & x_k & x_{k+1}\\ y_0 & y_{-1} & ... & y_{-k} & y_{-k-1}\end{array}\right).$$

If  $\binom{\bar x}{\alpha(\bar x)}$ approaches from the left side, then it has probability $L_{\alpha(x_0)\alpha(\bar x)}=L_{\bar x x_0}$ of to be accepted and to be the next nucleotide in the sequence. If it is accepted, then it is denoted as $\binom{x_{-1}}{y_1}$ and the sequence becomes $$\left(\begin{array}{ccccc} x_{-1} & x_0 & x_1 & ... & x_k\\ y_{1} & y_{0} & y_{-1} & ... & y_{-k}\end{array}\right).$$

If the nucleotide $\bar x$ is not accepted, then the original sequence does not change.\\

\item[step 4] Return to Step 0 setting the sequence obtained in Step 3 as the initial sequence.

\end{description}

\begin{figure}
\centering
\includegraphics[width=.8\linewidth=1.0]{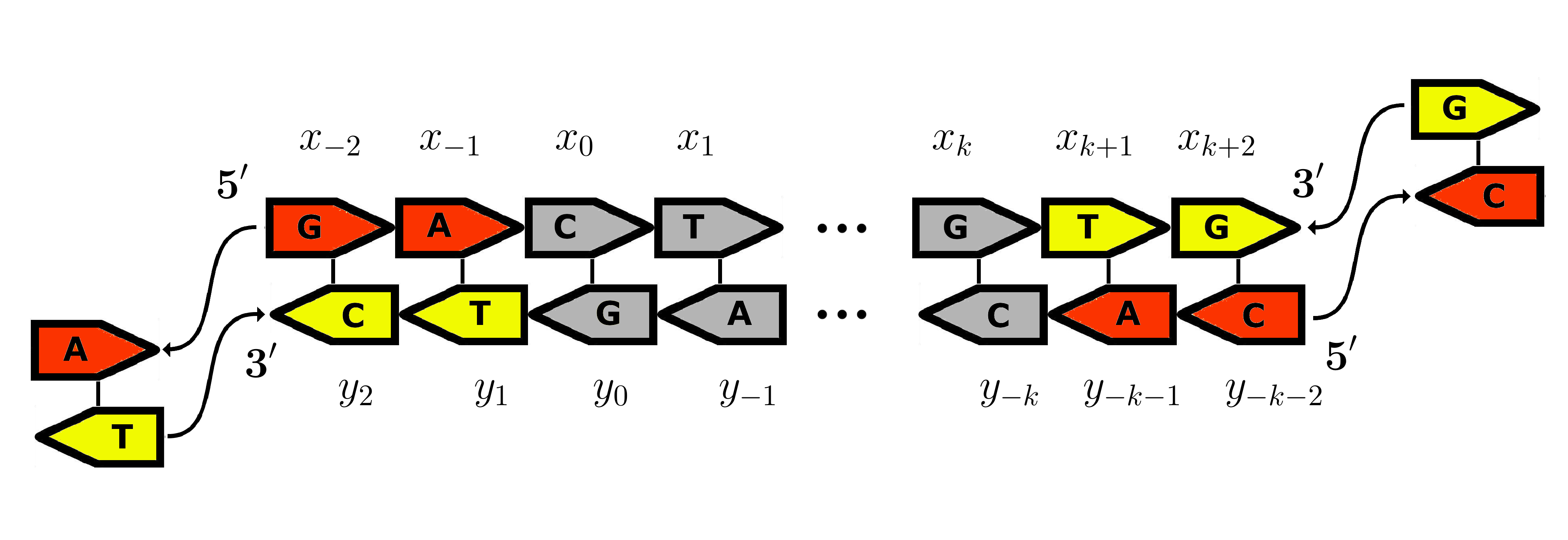}
\caption{A schematic presentation of the S-H model for constructing DNA sequences from an initial sequence $x_0x_1...x_k$ (depicted in gray). The construction sense is from $5'$ to $3'$ (from the left to the right in the top strand and from the right to the left in the bottom strand). A new nucleotide of type $G$ is selected with probability $M_G$ and
is appended to the end of the top strand with probability $L_{GG}$ (which corresponds to a nucleotide of type $C$ being selected with probability $M_C$ and appended to the start of the bottom strand with probability $L_{CC}$), while a
nucleotide of type $T$ is selected with probability $M_T$ and will be
attached to the end of the bottom strand with probability
$L_{CT}$ (which corresponds to a nucleotide of type $A$ being selected with probability $M_A$ and appended to the start of the top strand with probability $L_{AG}$). Each strand at the final DNA sequence obtained is the concatenation
of two Markovian processes: one depicted in yellow, whose estimated transition matrix is $P$ given by Equation \eqref{processP}; and its reverse complement depicted in red.}\label{S-H shortconstruction}
\end{figure}

Observe that after several (hundred of thousands or even millions) repetitions of the above process, one obtains a sequence duplex
$$\left(\begin{array}{ccccccccc} x_{-m} & ... & x_{-1} & x_0 & x_1 & ... & x_k & ... & x_n \\ y_m & ... & y_{1} & y_{0} & y_{-1} & ... & y_{-k} & ... & y_{-n}\end{array}\right),$$
where $m,n>0$ are two close numbers. Furthermore, observe that since $M_j=M_{\alpha(j)}$ and $L_{ij}=L_{\alpha(j)\alpha(i)}$, the process adds nucleotides of type $j$ to the right size of the Watson-strand (3') with the same probability that it adds nucleotides of type $j$ to the left size of the Crick-strand (3'). Therefore, it follows that the semi strands $(x_\ell)_{k+1\leq \ell\leq n}$ and  $(y_\ell)_{0\leq \ell\leq m}$ have equal oligonucleotide distribution (and then also the semi strands $(x_\ell)_{-m\leq \ell\leq 0}$ and  $(y_\ell)_{-n\leq \ell\leq -k-1}$ share a common oligonucleotide distribution). Hence, if we suppose that the initial sequence used in Step 0 is not so large (relatively to the final length of the genome), then by assessing the frequencies of oligonucleotides in $(x_\ell)_{0\leq \ell\leq n}$ and $(y_\ell)_{0\leq \ell\leq m}$ we would find that both have similar oligonucleotide distributions, while their complementary parts would also share a common distribution of oligonucleotides. Furthermore, it follows that each of the strands presents ISP for oligonucleotides and, moreover, if neither $\A$ nor $M$ are too uniform, then each strand will be composed for two chirochores with approximately the same length.

Note that the properties imposed on the matrix $\A$ and on the probability vector $M$ come from the Watson-Crick pairing and from the fact that the process assumes {\em no-strand-bias conditions} together a {\em directional-bias condition} (the DNA grows from 5' to 3'). 
In fact, due to the Watson-Crick pairing, the probability of a nucleotide of type $j$ to be available as candidate for being added to the sequence is equal to the probability of nucleotide of type $\alpha(j)$ to be available too, which means $M_j=M_{\alpha(j)}$. 
Furthermore, the probability of a nucleotide of type $j$ to be attached to the 3' extremity of a sequence where there is a nucleotide of type $i$ is given for $L_{ij}$, but looking in the complementary strand this is the same chance of the nucleotide of type $\alpha(i)$ to be attached in the the 3' extremity of the nucleotide $\alpha(j)$, and therefore $L_{ij}=L_{\alpha(j)\alpha(i)}$.

We remark that, in spite of each strand in the sequence $\binom{x_{\ell}}{y_{-\ell}}_{-m\leq\ell\leq n}$ produced by the S-H model to be a concatenation of two Markovian sequences and the fact that actual genomes are not Markovian \cite{Fukushima_et_al2000,Li_92,Yu_et_al2000}, first order Markov chains may well capture features of the nucleotide distributions in actual genomes \cite{Gao_et_al09}.

Now, suppose that we have a large sequence $(z_t)_{0\leq t\leq K}$ which corresponds to the Watson strand of a sequence that was produced according to the S-H model, and also suppose we do not know neither $M$ nor $\A$ that were used to produce such sequence nor the point where the the first chirochore ends and the second chirochore starts. We can first try to determine such point using some technique as proposed in \cite{FrankLobry00,Lobry96-2,Mackiewicz04}, and so, to denote the sequence in the form $(x_\ell)_{-m\leq \ell\leq n}$. Hence, we can count nucleotides and dinucleotides in $(x_\ell)_{0\leq \ell\leq n}$ and to estimate the Markovian measure $(P,\pi)$ of the sequence, where $\pi=(\pi_i)_{i=A,C,G,T}$ and $P=(P_{ij})_{i,j=A,C,G,T}$ (if $(x_\ell)_{0\leq \ell\leq n}$ is sufficiently large, then we can suppose that the process is near of the equilibrium and $(P,\pi)$ is a stationary Markov chain). It follows that $P$ is such that
\begin{equation}\label{processP}
P_{ij}:=\frac{L_{ij}M_j}{\displaystyle\sum_{k=A,C,G,T} L_{ik}M_k},
\qquad  \forall i,j=A,C,G,T.
\end{equation}

Hence, for $m$ and $n$ sufficiently large the Markov measure $(W,\omega)$ given by 

\begin{equation}\label{concatenatedW}
\omega_i:= t\pi_i+(1-t)\pi_{\alpha(i)}\qquad and\qquad 
W_{ij}:=\frac{t\pi_iP_{ij}+(1-t)\pi_{\alpha(j)}P_{\alpha(j)\alpha(i)}}{\omega_i},
\end{equation}
with $t=(n+1)/(m+n+1)\approx 1/2$,  will approximate the Markov measure estimated for the whole strand $(x_\ell)_{-m\leq \ell\leq n}$  by counting its mononucleotides and dinucleotides (we enforce that $(x_\ell)_{-m\leq \ell\leq n}$ is not in general a Markovian sequence, since it is the concatenation of two Markovian sequences).

In other words, the S-H model considers that a bacterial DNA sequence is a particular realization of a process which is as random as it is possible (and so has the maximum possible entropy) given the constraints imposed by: The availability of new nucleotides; The probabilities of each nucleotide being accepted in the sequence; The fact that new nucleotides are added only at the extremities of the strands. Thus the entropy of a DNA strand $X=(x_\ell)_{-m\leq \ell\leq n}$ would be function of $\A$, $M$ and $0<t<1$ and for $m$ and $n$ large it can be approximated by
$$h(X)=-\sum_{i,j=A,C,G,T}\omega_iW_{ij}\log(W_{ij}),$$
where $(W,\omega)$ is the Markov measure given in \eqref{concatenatedW}.

A matrix $P$ satisfying \eqref{processP} for some $\A$ and $M$ with $L_{ij}=L_{\alpha(j)\alpha(i)}$ and $M_i=M_{\alpha(i)}$ is said to be an $\aleph$-generated matrix and a sequence which is a realization of the stationary Markov chain $(P,\pi)$ is said to be an $\aleph$-generated sequence. In \cite{HartSobottka19} it was given formulations of the S-H model as a couple of two hidden Markov chains \cite[Theorems 1 and 2]{HartSobottka19}. Furthermore, \cite[Theorems 4]{HartSobottka19} provided a way to find a matrix $\A$ and a vector $M$ that generated a given $\aleph$-generated matrix $P$, while \cite[Theorems 11 and 12]{HartSobottka19}
proved that in a sequence $\binom{x_{\ell}}{y_{-\ell}}_{-m\leq\ell\leq n}$ produced by the S-H model we have all the following sequences being $\aleph$-generated:
$$(x_\ell)_{-m\leq\ell\leq n}\qquad
(x_\ell)_{-m\leq\ell\leq 0}\qquad
(x_\ell)_{0\leq\ell\leq n}\qquad
(y_\ell)_{-n\leq\ell\leq m}\qquad
(y_\ell)_{-n\leq\ell\leq 0}\qquad
(y_\ell)_{0\leq\ell \leq m}.$$
In particular, \cite[Theorem 12]{HartSobottka19} proved that sequences presenting ISP are always $\aleph$-generated. Finally it was proposed two measures to asses how far some given sequence is from being $\aleph$-generated and what is the matrix $\A$ that best fits the given sequence. 

In \cite{SobottkaHart11} it was analyzed 1049 bacterial genomes looking for the matrices $\A$ and vectors $M$ which produce the Markov measures given by Equation \eqref{concatenatedW} that best fits the Markov measure estimated from each genome. Then it was determined a matrix $\bar\A$ by taking the average of all the found matrices $\A$. This matrix $\bar\A$ was used to produce several stationary Markov chains $\{(W(s),\omega(s))\}_{s\in I}$, where each $W(s)$ was produced using $\bar\A$ according \eqref{concatenatedW}, and by taking $M=(s,0.5-s,0.5-s,s)$ for distinct values of $0<s<0.5$. Therefore, the relationship between their mononucleotide and dinucleotide frequencies were compared with the respective relationships found in the actual genomes (see \cite[Figure 4]{SobottkaHart10} and \cite[Fig. 2]{SobottkaHart11}). 

Since the relationship between mononucleotide and dinucleotide frequencies revealed patterns in the set of genomes which are apparently not related to specific DNA functions, and Markovian measures produced by the S-H model using a same 
matrix $\bar\A$ seem to fit to those patterns, then the authors postulated that: $(i)$ Any bacterial genome might be a realization of the S-H model by using a same matrix $\bar\A$ and varying $M$; The other features that arise through the model, namely ISP and SCA, might also be conserved patterns of a primordial genome (or their analogous in the pre DNA world or in the prebiotic world).

An intuitive interpretation of the existence of a unique matrix $\bar\A$ could be that the probability of a new nucleotide of type $j$ to be added to the sequence where there is a nucleotide of type $i$ could depend mainly on chemical
and other physical properties of these nucleotide bases, while $M$ would express some availability
of each nucleotide type in the environment. However, a biological interpretation of the S-H model remains open. 

Furthermore, the occurrence of ISP and SCA  as conserved patterns, should also deal with the problem of how these features have been conserved in modern genomes. In \cite{SobottkaHart11} it was speculated that relatively few uniformly distributed mutations on the primitive genome do not affect the occurrence of these features. In particular, since coding regions are, in general, sensible to mutations by loosing their functions, if some feature of a primordial genome was preserved in an actual genome, then it was likely preserved in coding regions (it is known as the neutral theory for molecular evolution \cite{Hey99,Kimura83}). Such assumption is consistent with the fact that SCA is directly observable in genomes where
a large proportion of their sequences correspond to coding regions, and also in the other genomes when one discharges non-coding regions \cite{Poptsova_et_al09}. It is also consistent with the observation made in \cite{LobrySueoka02} by Lobry and Sueoka that some portions of the genome
 are not free to deviate from ISP due to selective pressures.

Evolutive models for ISP and SCA may also lie in the neutral theory \cite{GaltierLobry97,Lobry95} by considering that the mutations that leads to these distributional features have not affected the fitness of the species\footnote{There are also {\em selectionist} perspectives for the emergence of ISP and SCA in bacterial genomes (see \cite{Forsdyke21}).}. However, while neutral evolutive models appeal to the neutral theory to explain how ISP and SCA emerged in bacterial genomes, the claim of  ISP and SCA  as conserved patterns appeal to the neutral theory to explain why these features did not disappear from bacterial genomes.\\

To finish this section, we highlight that other models were successful in proving that ISP can be obtained as direct consequence of the randomness under the constraints imposed by the Watson-Crick pairing of the duplex, without appealing for biological processes. For instance, Hart, Martínez and Olmos proposed an elegant  statistical-mechanics based model for ISP \cite{Hart_et_Al12}. More specifically, in their work it was proved that the Gibbs distribution associated with the chemical energy between the Watson and Crick strands satisfies ISP, and so ISP would be a probabilistic consequence of Watson-Crick pairing. Later in \cite{Fariselli21} Fariselli et al. revisited these ideas and formulated the problem in terms of informational entropy. Although these formulations do not address the problem of how SCA emerges in bacterial DNA together ISP, they could efficiently explain ISP for general double-stranded genomes. In fact, such statistical mechanic approaches provide a framework to argue that evolutive pressures and several random mutations could erase SCA-like features from genomes while pushing ISP.

\section{The S-H model as an evolutive model}

The S-H model could also have an evolutive interpretation. In fact, instead to interpret the model as the formation of a primitive genome, it could be interpreted as the process throughout a primitive genome (already capable of replication and then under evolutive pressures) have added bases along the time. More specifically, we could suppose that each bacterium evolved from some small-genome organism with two replichores (suppose that the initial sequence given in \eqref{eq:primitiveDNA} is the genome of such organism). Therefore, along the time new mononucletoide bases or entire oligonucleotides could have been added to its replichores near of the replication terminus. The process would work as originally proposed in \cite{SobottkaHart11} and even if the primitive genome did not present ISP and SCA, the final genome would present both features. 

This interpretation of the S-H model as an evolutive model could be supported by the fact that in actual bacterial genomes the chirochores almost coincide with the replichores. In fact, if the the S-H model account for the origin of a primordial genome, the ability to replicate would appear in some moment of sequence formation described by the S-H model, but it would not necessarily occur in the first stages of the sequence and then chirochores and replichores would not coincide. Hence, by considering that the original sequence used to start the construction according to the S-H model was already able to replicate the S-H model could be accounting for mutations as (oligo)nucloetide additions, substitutions, inverted transpositions and inverted duplications, that occurred predominantly  
at the end of each replichore. In particular inverted transposition and inverted duplications should occur by predominantly transposing or copying an oligonucleotide from the end of a leading strand in one replichore to end of the leading strand  in the other replichore. The probability for one of these events could be given by the chance of such error occurs, the probability of a given oligonucleotide to be available in the environment and/or to be present in one of the strands (which is related to the vector $M$), and the chance of the copy of the oligonucleotide to be accepted in the other strand (which is related to the matrix $\A$). This process would be similar to the inverted transpositions proposed in  \cite{AlbrechtBuehler2006} but with the difference that it is possible that each strand finishes with a copy of the same oligonucleotide and it would occur predominantly at the end of each replichore. 

 It is interesting to highlight that such mechanism would also be consistent with the fact that it has been observed that leading strands of of bacterial genomes usually present genes coding for the same functions in both replichores and at the same distance of the replication origin or replication terminus \cite{Eisen_et_Al_00,Mackiewicz01}. 
 
 Seen as evolutive, the S-H model might not longer be neutralist even still considering that it is randomly guided by some $\A$ and $M$. This because it would be very possible that actual genomes are those whose inverted duplications increase their fitness with a copy of genes in both leading strands. If this, then the matrix $\A$ estimated from some actual genome could not only be shaped by chemical constraints for each nucleotide base accepting other nucleotide in the sequence, but also by the fact that actual genomes are those for which the addition of new (oligo)nucleotides improved their fitness.

\section{Final discussion}

Intra-strand parity has been studied since it was observed by Chargaff \cite{Chargaff} in 1951. On the other hand, strand compositional asymmetry has been extensively studied for the last 25 years \cite{ Freeman_et_al98,Grigoriev98,Lobry96-1}.
In this article we have discussed some models for ISP and SCA in bacteria genomes, by focusing on two different perspectives to explain these features: Evolutive models that try to explain ISP and SCA as consequence of mutations; Conservative models which argue that ISP and SCA are features from primitive genomes that have been conserved through out the genome evolution. 
It is interesting to highlight that all the examined models derive ISP and/or SCA  from the Watson-Crick pairing of the DNA double helix. In particular, it is not naive to assume that the Watson-Crick pairing is influencing ISP since ISP holds almost exclusively for double-stranded DNA sequences \cite{MitchellBridge}. 

As we have seen, ISP and SCA in bacterial genomes are complementary features, since ISP only emerges in the whole strands as consequence of the fact that SCA means that the strand has two chirochores where one neutralizes the nucleotide skews of the other. Therefore, it is reasonable to expect that any explanation for ISP and SCA should address both features simultaneously or, at least, to expect that the mechanism behind one of these features is compatible with the mechanism leading to the other feature, in the sense that both could occur without one neutralizing the effects of the other.

The S-H model \cite{SobottkaHart11} captured several features of the bacterial genomes, including ISP and SCA and a new underlying pattern shared by bacterial genomes.
However, although recent works have addressed how single and double stranded RNA could emerge and evolve in a random pool of oligonucleotides  \cite{Derr_et_al12,Ianeselli22,Kudella21}, it lacks a plausible interpretation for the S-H model. In particular, the widely accepted hypothesis of the RNA world that would precede\footnote{Recently, it have been proposed that RNA and DNA have co-evolved (see \cite{LeVayMutschler19}).} the emergence of  DNA genomes \cite{Rich62,RobertsonJoyce12,Walter86} leads to suppose that any interpretation for the S-H model as the process underlying the formation of primitive genomes should possibly be based on evidences of as RNA 
or even other pre-RNA candidates polymerize \cite{EngelhartHud10,Mizuuchi22,Yashima08}. In 
particular, it would be necessary more research on how double-stranded polymers polymerizes \cite{Benner_et_Al10,Yashima08,Zhang21}. 

Among the evolutive models we have analyzed we can derive ISP and SCA from the substitutions presented in \cite{FrankLobry99} whenever, besides deamination, the leading strands  also are prone to  depyrimidination. On the other hand, from an initial genome with replichores having distinct distributions of nucleotides (but not necessarily with SCA), the inverted transpositions proposed by Albrecht-Buehler \cite{AlbrechtBuehler2006}, but occurring predominantly by changing oligonucleotides from one replichore to the other, would lead for ISP and partial SCA, since they could not explain alone neither the keto-amino skew nor the purine-pyrimidine skew in the leading strands.

Furthermore, if besides inverted transpositions we consider the possibility of inverted duplications, and both occurring predominantly in the extremities of the replichores we get a formulation of the S-H model as an evolutive model that could explain: ISP, SCA, why replichores and chirochores almost coincide, the pattern of the gene locations in bacterial genomes \cite{Eisen_et_Al_00,Mackiewicz01}, and why the Markov measures estimated from the bacterial genomes seem to accomplish \eqref{concatenatedW} for some $\A$ and $M$ \cite{SobottkaHart11}. Note that, in the S-H model the keto-amino and purine-pyrimidine skews in the leading strand are captured by the particular form of the matrix $\A$, and  could be interpreted as originated from chemical constraints or evolutive pressures.

The debate about the origins of ISP and SCA in bacteria is far from the end, and even when considering only evolutive models (which are the most accepted explanations at the moment) the debate between the neutralism and selectionism remains heated (see \cite{Forsdyke21} for criticism to the neutralist perspective).  More research is need to test all the proposed models and to obtain more evidences that allow to discharge or to validate them. 

In particular, the allegation of the S-H model that each chirochore could be approximated for an $\aleph$-generated matrix should still be properly investigated. The estimations made in \cite{SobottkaHart11} were crude and could be improved by using the results obtained in \cite{HartSobottka19}. Furthermore, from \cite[Equations (4) and (5)]{HartSobottka19}, if the hypothesis that each bacterial genome is a particular realization of the S-H for a same matrix $\bar\A$ is true, then we should find that the following ratios
$$ 1)\ \frac{f_{AA}}{f_{AT}}\qquad  2)\ \frac{f_{AC}}{f_{AG}}\qquad 3)\ \frac{f_{CA}}{f_{CT}}\qquad 4)\ \frac{f_{CC}}{f_{CG}}\qquad 5)\ \frac{f_{GA}}{f_{GT}}\qquad 6)\ \frac{f_{GC}}{f_{GG}}\qquad 7)\ \frac{f_{TA}}{f_{TT}}\qquad 8)\ \frac{f_{TC}}{f_{TG}},$$
where $f_{ij}$ denotes the frequency of dinucleotide $ij$ in the leading (equivalently, lagging) strand, do not vary significantly for distinct bacterial genomes.

\section*{Acknowledgments}

\noindent 
The author thanks Eduardo da Veiga Beltrame for reading an early version of this manuscript and making valuable suggestions.


\end{document}